\documentclass[a4paper]{article}
\usepackage[final]{graphicx}
\newcommand{\ksl}{/\kern-6pt k}
\newcommand{\qsl}{/\kern-6pt q}

\title{Invariances of regularized QED diagrams}
\author{Marijan Ribari\v c and Luka \v Su\v ster\v si\v c\thanks{Corresponding author. Phone +386 1 477 3258; fax +386 1 423 1569; electronic address: \tt luka.sustersic@ijs.si\rm} \\Jo\v zef Stefan Institute, p.p.3000, 1001 Ljubljana, Slovenia }
\date{}

\begin{document}

\maketitle

\begin{abstract}
We consider properties of connected diagrams with fermion-photon interaction and such fermion and photon propagators and vertex function that the values of these diagrams are finite. We establish the properties of these propagators and vertex function that imply that these diagrams are invariant under C, P, T, CP, CT, PT, or CPT transformations up to some phase factor common to each process. We introduce eight new transformations related to Hermiticity and establish the conditions under which they leave the tree transition probabilities invariant. We determine such general Lorentz form-invariant fermion and photon propagators and fermion-photon vertex functions that make diagrams Lorentz-invariant.
\end{abstract}

\section{Introduction}

In QED, perturbative calculation of transition amplitudes results in Feynman diagrams where integrals over independent loops diverge. To make mathematical manipulations on these diagrams possible, various regularization methods can be used to make all such integrals finite. Regularization amounts to a temporary modification of the original definition of QED: the parameters introduced by regularization are subsequently removed by a renormalization scheme. 

It is a sixty year old open question whether there is a regularization method such that we can regard the regularized Feynman diagrams as part of a plausible physical theory, and the parameters introduced by regularization as measurable, physical quantities. The motivation of this work is the following belief: \it There exists a physically plausible theory that can provide the transition amplitude for each quantum-electrodynamic scattering process in the form of an asymptotic series in powers of the electron  charge such that: (a)~Terms of this series can be described by finite-valued, connected Feynman diagrams with fermion-photon interaction. (b)~This series implies the present results of QED as a low-energy approximation. \rm  This series may be non-renomalizable, e.g. it may contain diagrams due to Pauli's term; for some related comments see Lapage\cite{Lapage}, Sect.~3.1; Collins\cite{Collins}, Sect.~6.4; and Weinberg\cite{Weinberg}, Sect.~12.3. 

On this belief, we will study invariances of momentum-space, connected Feynman diagrams of QED with massive photons \cite{Hooft,Veltman} that we regularize as follows:
\begin{itemize}
\item[(A)] We replace the fermion and photon propagators of QED with generic ones: two $4 \times 4$ matrix-valued functions, say $P_f(k)$ and $P(k)$ respectively. We assume that
\begin{equation}
     \tilde{P}(k) = P(-k) \,,
     \label{Pppro}
\end{equation}
where $\tilde{ }$ denotes transpose; a photon propagator has this property if it is determined by a Lagrangian for real four-vector fields, see\cite{Hooft}, eq.(2.5). 
\item[(B)] We replace the QED vertex factor $\gamma^\mu$ with a generic vertex function $\Gamma(k',k'')$ for the fermion-photon interaction, whose four-vector components $\Gamma_\mu(k',k'')$ are $4 \times 4 $ matrix-valued functions of the two fermion momenta.
\item[(C)] We assume that the fermion and photon propagators $P_f(k)$ and $P(k)$ and the fermin-photon vertex function $\Gamma(k', k'')$ regularize QED diagrams, i.e. that in each so regularized QED diagram all integrals over independent loops converge.
\end{itemize}
We make no further assumption about $P_f(k)$, $P(k)$ and $\Gamma(k', k'')$; in particular, we do not require that the Feynman series of such diagrams is renormalizable. \it Henceforth all diagrams in this paper are such as just specified. \rm

In the following section, we specify such graphical and algebraic representations of diagrams that are convenient for our considerations. In Section~\ref{secconservation} we point out that diagrams conserve total charge and momentum. In Section~\ref{secsymmetries} we give the properties of regularizing fermion and photon propagators and of vertex function that suffice to make the value of each diagram invariant under C, P, T, CP, CT, PT, or CPT transformation up to a phase factor that is common to all diagrams of the same process; QED propagators and vertex function have these properties. In Section~\ref{secinvariances}, we introduce a new, Hermiticity-related transformation H of diagrams. We give properties of the regularizing fermion and photon propagators and of the vertex function that suffice to make the values of all lowest-order diagrams of the same process invariant under H, HC, HP, HT, HCP, HCT, HPT, or HCPT transformation up to a common phase factor; again, QED propagators with $i\epsilon = 0$ and vertex function have these properties. In this paper, we define all fifteen discrete transformations considered (C, P, T, H, and their products) as \it active transformations of scattering processes. \rm In Section~\ref{secgeneral}, we establish the general fermion propagator, photon propagator, and vertex function that make diagrams Lorentz invariant; thereby we obtain a special case of the CPT theorem. In Section~\ref{secproperties}, we apply the results of Sections~\ref{secsymmetries}--\ref{secgeneral} to obtain conditions under which transition probabilities for quantum-electrodynamic processes (i)~are equal to all orders for two processes related by an active C, P, T, CP, CT, PT, or CPT transformation or a passive Lorentz transformation, and (ii)~vanish to all orders for processes invariant under one of these transformations. In addition, we give conditions under which the lowest-order approximations to transition probabilities are equal for two processes related by an active H, HC, HP, HT, HCP, HCT, HPT, or HCPT transformation. In Section~\ref{secformal}, we consider chiral and two additional formal transformations under which transition probabilities are invariant. In Section~\ref{secconclusions}, we close with some comments about the significance of the obtained results.

\begin{figure}[t]
\begin{center}
\includegraphics[bb= 160 420 450 720]{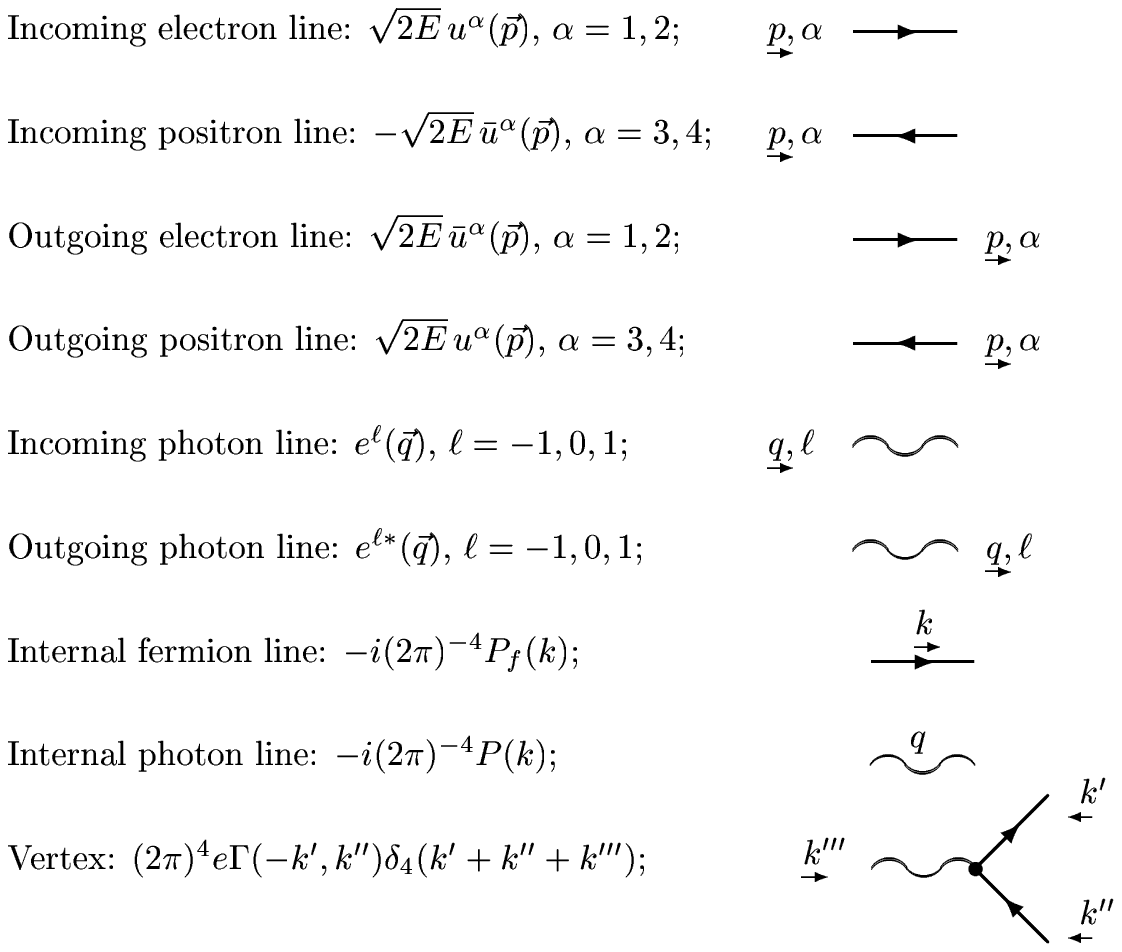}
\caption[prva slika]{Notation for elements of Feynman diagrams}
\end{center}
\label{elementi}
\end{figure}

Regarding metric, terminology, and other conventions, we follow t'Hooft and Veltman \cite{Hooft,Veltman}. In particular, $\mu, \nu = 1,2,3,4$; $k = (\vec{k}, k_4)$, $k_4 = ik_0$; $k^2 = \vec{k}^2 + k_4^2 = \vec{k}^2 - k_0^2$; the Dirac matrices $\gamma^\mu$ are such that $\gamma^{\mu\dagger} = \gamma^\mu$, $\tilde{\gamma}^\mu = (-1)^\mu \gamma^\mu$, $\gamma^\mu \gamma^\nu + \gamma^\nu\gamma^\mu = 2\delta_{\mu\nu} I$, and $\gamma^5 = \gamma^1 \gamma^2 \gamma^3 \gamma^4$. Elements of diagrams are denoted as specified in Fig.~\ref{elementi}: External line factors $u^\alpha(\vec{p})$ and $e^\ell(\vec{q})$ for fermions and circularly polarized photons are given in [4], App.~A; $\bar{u}(\vec{p}) = u^\dagger (\vec{p})\gamma^4$; $p = (\vec{p}, iE)$ with $E = + \sqrt{ \vec{p}^2 + m^2}$, and $q = (\vec{q}, iq_0)$ with $q_0 = +\sqrt{\vec{q}^2 + \kappa^2}$ are the momenta of an external fermion and photon, where $m$ is the electron mass and $\kappa\ll m$ is the photon mass; spin number $\alpha$ equals $1$ [$4$] for a spin-up electron [positron], and $2$ [$3$] for a spin-down electron [positron]; spin number $\ell$ equals $1$ [$-1$] for a spin-up [spin-down] photon, and $0$ for a spin-$z$ component zero photon; and momenta $p$, $q$, $k$, $k'$, $k''$, and $k'''$ flow in the directions of the underlying arrows; $e$ is the electron charge.

\section{Regularized QED diagrams}
\label{secdiagrams}

To calculate the perturbative transition probability for a quantum-electrody\-na\-mic scattering process, we need the corresponding set of diagrams. Only the set of all external lines of a diagram is determined by the process under study; and it uniquely identifies this process. For diagrams of the same process, the number of their vertices
\begin{equation}
    V = V_{{\rm min}},\; V_{{\rm min}} + 2, \ldots, 
        \quad V_{{\rm min}} = E_p + \max (E_f - 2, 0) \,,
    \label{Vmin}
\end{equation}
and the number of their internal lines
\begin{equation}
    I = (3V - E_f - E_p)/2 = I_{{\rm min}}, \; I_{{\rm min}} + 3, \ldots, 
        \quad I_{{\rm min}} = V_{{\rm min}} - \min (E_f, 1) \,,
    \label{Imin}
\end{equation}
where $E_f$ [$E_p$] are the number of external fermion [photon] lines.

We will denote by $N_+$ [$N_-$] the total number of positrons [electrons] of a process, and by $S_+$ [$S_-$] the total number of spin-up [spin-down] fermions. Since $N_+ + N_- = S_+ + S_- = E_f$, and $E_f$ is even, we have
\begin{equation}
     (-1)^{N_+} = (-1)^{N_-} \,, \qquad (-1)^{S_+} = (-1)^{S_-} \,.
     \label{nspm}
\end{equation}

\begin{figure}[t] 
\begin{center}
\includegraphics[bb=250 550 400 700]{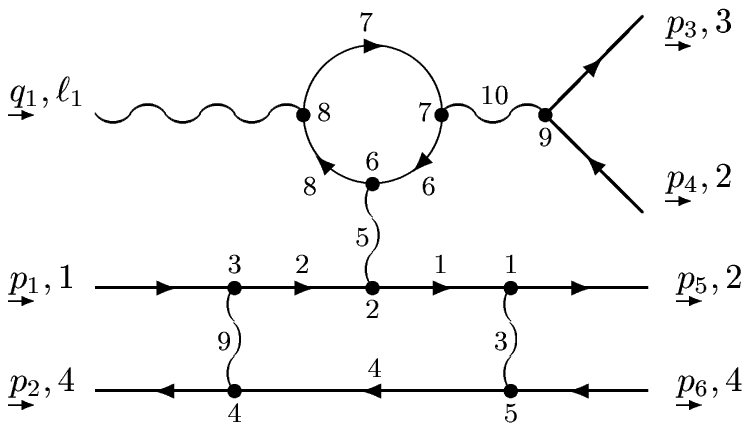}
\caption[druga slika]{An example of a graph of a process where the number of external fermion lines $E_f= 2O_f =6$, the number of external photon lines $E_p=1$, the minimal number of vertices $V_{{\rm min}}= 5$, the minimal number of internal lines $I_{{\rm min}}= 4$, the number of positrons [electrons] $N_+ = 3$ [$N_- = 3$], the difference between the number of incoming [outgoing] positrons and electrons $D_{in} = 0 $ [$D_{out} = 0$], and the number of spin-up [spin-down] fermions $S_+ = 3$ [$S_- = 3$]. This graph has: $3$ open fermion lines ($O_f =3$), $1$ closed fermion loop ($L_f=1$), $9$ vertices ($V=9$), $6$ internal fermion lines ($I_f = 6$), $4$ internal photon lines ($I_p =4$), and $2$ independent loops ($L_i =2$). We numbered the vertices and internal lines by an extended 't Hooft and Veltman method.}
\end{center}
\label{graf}
\end{figure}

It will be convenient to graphically represent a diagram by the corresponding \it graph. \rm We draw the incoming lines on the left-hand side, and the outcoming ones on the right-hand side of the graph, with lettering denoting the definite momenta and spin numbers of electrons, positrons and photons of the process; for an example see Fig.~\ref{graf}.

We \it algebraically represent \rm the value of each diagram as follows:

Let us take a diagram that has $O_f \ge 0$ open fermion lines, $E_f = 2O_f$ external fermion lines, $L_f \ge 0$ closed fermion loops, $E_p \ge 0$ external photon lines, $V \ge 1$ vertices, $I \ge 0$ internal lines, $I_f = V - O_f \ge 0$ internal fermion lines, and $I_p = (V- E_p)/2 \ge 0$ internal photon lines.
\begin{itemize}
\item{} We denote by $p_j$, $E_j$, and $\alpha_j$ the momentum, energy, and spin number of the external fermion line $j$, $j  = 1$, $2$, \ldots, $E_f$, and by $q_n$ and $\ell_n$ the momentum and spin number of the external photon line $n$, $n = 1$, $2$, \ldots, $E_p$.
\item{} Following 't Hooft and Veltman\cite{Hooft}, App.~A, we number the vertices from $1$ to $V$ consecutively along the open fermion lines and closed fermion loops in the direction opposite the arrows on these lines; for an example see Fig.~\ref{graf}.
\item{} We number each of the $I_f$ internal fermion lines by the number of the vertex that the arrow on this line points to, and use the remaining numbers between $1$ and $I$ for the $I_p$ internal photon lines; which is possible since $I_f + I_p \ge V-1$, by (\ref{Vmin}) and (\ref{Imin}). 
\item{} We denote by $k_m$, $m = 1$, $2$, \ldots, $I$, the internal momentum that flows along the internal line $m$ out of the vertex with the higher number into the vertex with the lower number.
\end{itemize}
The \it value \rm of this diagram is the "product" of the following factors:
\begin{itemize} 
\item The numerical factor 
\begin{equation}
     (-1)^{L_f} i^{-I} e^V (2\pi)^{4(V-I)} 
     \label{Gvalue1}
\end{equation}
multiplied by a combinatorial factor, possibly with a minus sign to account for permutations of fermion lines relative to other diagrams.
\item $I$ Lorentz-invariant four-integrals over the internal momenta $k_m$, say
\begin{equation}
     \int d_4k_1 \cdots \int d_4k_I \,, \qquad d_4k_m = dk_{m0} dk_{m1} dk_{m2} dk_{m3} \,.
     \label{facint} 
\end{equation}

\item $V$ Dirac delta functions
\begin{equation}
     \delta_4( k'_1 + k''_1 + k'''_1 ) \cdots \delta_4( k'_V + k''_V + k'''_V ) \,,
     \label{delta1}
\end{equation}
where $k'_v$, $k''_v$ and $k'''_v$ are the momenta flowing into $v$-th vertex.
\item $O_f$ factors of open fermion lines. For an open fermion line with vertices numbered from $v$ to $v' >v$, and external momenta and spin numbers $p_j$ and $p_{j'}$ and $\alpha_j$ and $\alpha_{j'}$, we have the factor
\begin{eqnarray}
    \lefteqn{2\eta(\alpha_j) \sqrt{E_j E_{j'}}\, \bar{u}^{\alpha_j}(\vec{p}_j) \Gamma_{\mu_v} (\eta(\alpha_j) p_j, k_v) 
          P_f(k_v) \Gamma_{\mu_{v+1}} (k_v, k_{v+1})\cdots } \label{diag1} \\ && \kern10pt 
          \cdots \Gamma_{\mu_{v'-1}} (k_{v'-2}, k_{v'-1}) P_f(k_{v'-1}) \Gamma_{\mu_{v'}} (k_{v'-1}, 
          \eta(\alpha_{j'}) p_{j'}) u^{\alpha_{j'}}(\vec{p}_{j'}) \,, \kern20pt\nonumber  
\end{eqnarray}
where $\eta(\alpha) = {\rm sgn}(5 - 2\alpha)$; $\eta(\alpha_j) p_j$ [$\eta(\alpha_{j'}) p_{j'}$] is the momentum that flows out of [into] vertex $v_j$ [$v_{j'}$]. For an open fermion line with a single vertex, the corresponding factor is
\begin{equation}
    2\eta(\alpha_j) \sqrt{E_j E_{j'}}\, \bar{u}^{\alpha_j}(\vec{p}_j) \Gamma_{\mu_v} (\eta(\alpha_j) p_j, 
          \eta(\alpha_{j'}) p_{j'}) u^{\alpha_{j'}}(\vec{p}_{j'}) \,. 
    \label{diag3}
\end{equation}
\item $L_f$ factors of closed fermion loops. For a closed fermion loop with vertices numbered from $v$ to $v' > v$, we have the factor
\begin{equation}
     {\rm Tr} \left [ \Gamma_{\mu_v}(k_{v'}, k_v) P_f(k_v) \Gamma_{\mu_{v+1}}(k_{v}, k_{v+1})\cdots 
            \Gamma_{\mu_{v'}} (k_{v'-1}, k_{v'}) P_f(k_{v'} ) \right ] \,;
     \label{diag2}
\end{equation}
for a loop with a single vertex, the factor is
\begin{equation}
    {\rm Tr} \left [ \Gamma_{\mu_v} (k_v, k_v) P_f(k_{v}) \right ] \,. 
    \label{diag4}
\end{equation}
\item $I_p$ factors of internal photon lines. For the internal photon line $m$ connecting two vertices $v$ and $v' > v$, we have the factor
\begin{equation}
     P_{\mu_{v^{\vphantom{\prime}}} \mu_{v'}}(k_m) \,,
    \label{diag6}
\end{equation}
which equals $ P_{\mu_{v'} \mu_{v^{\vphantom{\prime}}}}(-k_m)$, by (\ref{Pppro}).
\item $E_p$ factors of external photon lines. For the incoming [outgoing] photon line $n$ connected to vertex $v$, we have the factor 
\begin{equation}
    e^{\ell_n}_{\mu_v}(\vec{q}_n) \qquad [e^{\ell_n *}_{ \mu_v}(\vec{q}_n)] \,.
    \label{diag5}
\end{equation}
\end{itemize}
We denote this algebraic representation, i.e. the "product" of factors (\ref{Gvalue1}), (\ref{facint}), \ldots, and (\ref{diag5}), by (\ref{Gvalue1})--(\ref{diag5}).

Henceforth we assume that the values of momenta and spin numbers of the process we consider are such that value (\ref{Gvalue1})--(\ref{diag5}) is well-defined and so finite; i.e. we will consider no process with exceptional momenta.

\section{Conservation laws}
\label{secconservation}

\subsection{Conservation of electric charge}

For each diagram, the total electric charge of incoming [outgoing] particles is proportional to the difference $D_{in}$ [$D_{out}$] between the number of incoming [outgoing] positron and electron lines. Now $D_{in} = D_{out}$, because (i)~$D_{in}$ [$D_{out}$] equals the difference between the number of left- and right-going open fermion lines on the left-hand [right-hand] side of the graph, and (ii)~each open fermion line starts either at one side of the graph and ends on the opposite side going in the same direction, or it ends on the same side of the graph going in the opposite direction, see Fig.~\ref{graf}. So for each diagram the total electric charges of incoming and outgoing particles are the same.

\subsection{Conservation of momentum}

The product (\ref{delta1}) of Dirac delta functions vanishes unless the external and internal momenta $p_j$, $q_n$, and $k_m$ satisfy $V$ balance equations
\begin{equation}
   k'_v + k''_v + k'''_v = 0, \qquad v = 1,2, \ldots V.
   \label{vercon}
\end{equation}
The sum of these equations reads
\begin{equation}
   \sum_{in} k_v - \sum_{out} k_v = 0 \,,
   \label{momcon}
\end{equation}
where $\sum_{in} k_v$ [$\sum_{out} k_v$] is the sum of the momenta of all incoming [outgoing] particles. So the value of each diagram vanishes unless the total momenta of incoming and outgoing particles are the same.

\it Independent loops. \rm When the external momenta $p_j$ and $q_n$ satisfy relation (\ref{momcon}), we can use relations (\ref{vercon}) to express $V-1$ of $I$ internal momenta $k_m$ in terms of the external momenta and the remaining $L_i = I - V + 1 \ge 0$ mutually independent internal ones. We can then rewrite factors (\ref{diag1})--(\ref{diag6}) in terms of the external momenta and the $L_i$ independent internal momenta, drop the $V-1$ four-integrals over dependent internal momenta, and replace the product (\ref{delta1}) of Dirac delta functions with the Dirac delta function $\delta_4(\sum_{in} k_v - \sum_{out} k_v) $. The remaining $L_i$ four-integrals over the independent internal momenta are finite by assumption.

From (\ref{Vmin}) we conclude that for diagrams of the same process, the number of their independent loops
\begin{equation}
     L_i = (V - E_f - E_p)/2 + 1 = 1 - \min(E_f, 1),\; 2 - \min(E_f, 1), \ldots ,
     \label{Lmin}
\end{equation}
and $L_i \ge L_f$. So there are no tree diagrams (where we can explicitly carry out all integrations (\ref{facint})) if $E_f = 0$, i.e. if the process involves only photons.

\section{Order-independent invariances}
\label{secsymmetries}

In this section, we consider seven active, discrete transformations that transform \it a scattering process and the set of its diagrams into another scattering process and the set of its diagrams \rm respectively. A transformation of a process is a change of its particles, which may include changing incoming [outgoing] particles into outgoing [incoming] ones, changing particles into antiparticles, reversing their three-momentum, and inverting their spin (see Table~\ref{drugatabla}). And a transformation of diagrams is defined by its action on graphs, which may include mirroring them, reversing the direction of arrows on fermion lines, reversing the direction of three-momentum on external lines, and changing the spin numbers (see Table~\ref{prvatabla}). We establish the properties of $P_f(k)$, $P(k)$ and $\Gamma(k',k'')$ that imply that the value of a diagram is invariant under these transformations up to a phase factor that is common to all diagrams of the same process. In Section~\ref{secproperties}, we will explore the implications of these results for transition probabilities.

\subsection{Charge conjugation}

By definition, \it charge conjugation, \rm C, changes each particle of a process into the corresponding antiparticle; and for each graph, it reverses the direction of all arrows on fermion lines, and replaces the fermion spin numbers $\alpha_j$ with $5-\alpha_j$. If
\begin{eqnarray}
      P_f(-k) &=& \gamma^4 \gamma^2 \tilde{P}_f(k) \gamma^2 \gamma^4 \,, \nonumber \\ 
     \Gamma_\mu( -k'', -k') &=& -\gamma^4 \gamma^2 \tilde{\Gamma}_\mu(k', k'') \gamma^2 \gamma^4 \,, \label{Ccond}
\end{eqnarray}
then the value of each diagram equals the value of the C-transformed diagram multiplied by the phase factor 
\begin{equation}
     (-1)^{E_p + N_+} \,,
     \label{CC3}
\end{equation}
common to all diagrams of the same process. This phase factor is consistent with charge-conjugation parities $+1$, $-1$, and $-1$ (or $-1$, $+1$, and $-1$, by (\ref{nspm}) ) of electron, positron, and photon respectively, see \cite{Weinberg}, Sec.~3.3.

To prove this, we write down the algebraic representation (\ref{Gvalue1})--(\ref{diag5}) for the value of the C-transformed diagram:
\begin{itemize}
\item[(i)] We number the vertices and lines, and define integration variables, as in the original graph. So the numerical factor (\ref{Gvalue1}), the integrals (\ref{facint}), the Dirac delta functions (\ref{delta1}), the internal photon line factors (\ref{diag6}), and the external photon line factors (\ref{diag5}) are the same as for the original diagram.
\item[(ii)]~The open-fermion-line factor (\ref{diag1}) for the C-transformed diagram reads 
\begin{eqnarray}
        \lefteqn{-2 \eta(\alpha_{j'}) \sqrt{E_j E_{j'} } \bar{u}^{5-\alpha_{j'}}(\vec{p}_{j'}) \Gamma_{\mu_{v'}}(
               -\eta(\alpha_{j'}) p_{j'}, -k_{v'-1}) P_f(-k_{v' -1}) \cdots } \label{CC2} \\ &&\kern90pt
        \cdots P_f(-k_v) \Gamma_{\mu_v} (-k_v, -\eta(\alpha_j) p_j) u^{5-\alpha_j}(\vec{p}_j)\,.\kern20pt \nonumber
\end{eqnarray}
Using (\ref{Ccond}),
\begin{eqnarray}
    u^{5-\alpha}(\vec{p}) &=& - \gamma^2  {u^\alpha}^*(\vec{p}) \,, \\
    \tilde{u}^\alpha \tilde{M} \gamma^4 u^{\beta *} &=& \bar{u}^\beta M u^\alpha \,, \label{CC5}
\end{eqnarray}
where $M$ denotes a $4\times 4$ matrix, we can show that (\ref{CC2}) is equal to the corresponding original factor (\ref{diag1}) multiplied by $ (-1)^{v'- v + 1} \eta(\alpha_j) \eta(\alpha_{j'})$.
\item[(iii)]~Similarly, the fermion-loop factor (\ref{diag2}) for the C-transformed diagram is 
\begin{equation}
     {\rm Tr} \left [ P_f(-k_{v'}) \Gamma_{\mu_{v'}} (-k_{v'}, - k_{v'-1}) \cdots 
           P_f(-k_v) \Gamma_{\mu_v} (-k_{v}, -k_{v'}) \right ] \,.
     \label{CC6}
\end{equation}
Using (\ref{Ccond}), $ {\rm Tr} \tilde{M} = {\rm Tr} M$ and ${\rm Tr} M_1 M_2 = {\rm Tr} M_2 M_1$, we can show that (\ref{CC6}) is equal to the factor (\ref{diag2}) for the original diagram multiplied by $(-1)^{v'- v + 1}$. 
\item[(iv)]~As the product of factors $(-1)^{v'- v + 1}$ of all open fermion lines and fermion loops equals $(-1)^V$, and \begin{equation}
     \Bigl( \prod_j \eta(\alpha_j) \Bigr) \Bigl( \prod_{j'} 
           \eta(\alpha_{j'}) \Bigr) = (-1)^{N_+} \qquad {\rm for} \qquad E_f > 0 \,,
     \label{produkt}
\end{equation}
the value of the C-transformed diagram equals the value of the original diagram up to the factor $(-1)^{V + N_+}$, which equals (\ref{CC3}) as $V = E_p + 2 I_p$.
\end{itemize}

\subsection{Space reflection}

By definition, \it space reflection, \rm P, reverses the direction of three-momentum of each particle; the P-transformed graph is obtained by replacing the momenta $p_j$ and $q_n$ of external lines with $R p_j$ and $R q_n$ respectively, where the $4\times 4$ matrix $R$ is such that $R k = (-\vec{k},ik_0)$, so that $R^2 = 1$. If 
\begin{eqnarray}
     P_f(k) &=& \gamma^4 P_f(Rk) \gamma^4 \,, \nonumber \\
     P(k)   &=& R P(Rk) R \,, \label{Pcond} \\  
     \Gamma_\mu(k', k'') &=& R_{\mu\nu} \gamma^4 \Gamma_\nu (Rk', Rk'') \gamma^4 \,, \nonumber
\end{eqnarray}
then the value of each diagram equals the value of the P-transformed diagram multiplied by the phase factor
\begin{equation}
     (-1)^{E_p + N_+} \,,
     \label{factor1}
\end{equation}
common to all diagrams of the same process. This phase factor is identical to the charge-conjugation phase factor (\ref{CC3}); it is consistent with intrinsic parities $+1$, $-1$, and $-1$ (or $-1$, $+1$, and $-1$, by (\ref{nspm}) ) of electron, positron and photon respectively, see \cite{Weinberg}, Sec.~3.3.

To check this statement we use relations
\begin{equation}
     u^{\alpha}(-\vec{p}) = \eta(\alpha) \gamma^4 u^\alpha(\vec{p}) \,,\qquad
     e^\ell_\mu(-\vec{q}) = - R_{\mu\nu} e^\ell_\nu(\vec{q}) \,,
     \label{ePrel}
\end{equation}
and take into account (\ref{produkt}) and that the Dirac delta function $\delta_4 (k)$ and the four-integral $\int d_4k$ are invariant under replacement of $k$ with $R k$.

\subsection{Time reversal}

By definition, \it time reversal, \rm T, changes each incoming [outgoing] particle into an outgoing [incoming] particle with reversed direction of three-momentum and inverted spin; the T-transformed graph is the mirror image of the original graph where we (a)~remirrored the lettering, (b)~reversed the direction of all arrows on fermion lines, (c)~replaced external-line momenta $p_j$ with $Rp_j$ and $q_k$ with $Rq_k$, (d)~replaced fermion spin numbers $\alpha_j$ with $\beta(\alpha_j)$, where 
\begin{equation}
   \beta(1) = 2 \,, \quad \beta(2) = 1 \,, \quad \beta(3) = 4 \,, \quad \beta(4) = 3 \,, \label{betadef}
\end{equation}
and (e)~replaced photon spin numbers $\ell_n$ with $-\ell_n$. If
\begin{eqnarray}
      P_f(k) &=& \gamma^2 \gamma^5 \tilde{P}_f(Rk) \gamma^5 \gamma^2 \,, \nonumber \\
      P(-k)  &=&  R P(R k) R \,, \label{Tcond} \\
      \Gamma_\mu(k'', k') &=& \gamma^2 \gamma^5 R_{\mu\nu} \tilde{\Gamma}_\nu (Rk', Rk'') \gamma^5 \gamma^2 \,, \nonumber
\end{eqnarray}
then the value of each diagram equals the value of the T-transformed diagram, multiplied by the phase factor
\begin{equation}
     (-1)^{E_p + N_+ + S_+ } \,,
     \label{TRpower}
\end{equation}
common to all diagrams of the same process, cf.~(\ref{nspm}).

To prove this assertion, we need relations (\ref{CC5}),
\begin{equation}
     \eta(\alpha) u^{\beta(\alpha)}(-\vec{p}) = -(-1)^\alpha \gamma^2 \gamma^4 \gamma^5 u^{\alpha\,*}(\vec{p}) \,, \qquad
     e^{-\ell*}_\mu (-\vec{q}) = - R_{\mu\nu} e^\ell_\nu(\vec{q}) \,,
     \label{TR2}
\end{equation}
\begin{equation}
     (-1)^{\sum_j \alpha_j + \sum_{j'} \alpha_{j'} } = (-1)^{N_+ + S_+ } \,.
     \label{TR1}
\end{equation}
We argue by analogy to the case of C and P transformations. We index the vertices and choose internal momenta in such a way that the factors (\ref{diag1})--(\ref{diag5}) for the T-transformed diagram equal the original factors (\ref{diag1})--(\ref{diag5}), with reversed order of multiplication in (\ref{diag1}) and (\ref{diag2}), and with $u^{\alpha}(\vec{p})$, $e^{\ell}(\vec{q})$, and $P(k)$ replaced with $\eta(\alpha) \bar{u}^{\beta(\alpha)}(-\vec{p})$, $e^{-\ell\,*}(-\vec{q})$, and $P(-k)$ respectively. Using (\ref{Tcond})--(\ref{TR1}) and (\ref{CC5}), and introducing new integration variables, we can prove the above result.

\subsection{CP transformation}

CP transformation is the result of the combined application of C and P transformations, taken in any order. The CP-transformed graph is obtained by (a)~reversing the direction of all arrows on fermion lines, (b)~replacing the external-line momenta $p_j$ with $Rp_j$ and $q_n$ with $Rq_n$, and (c)~replacing the spin numbers $\alpha_j$ with $5 - \alpha_j$. If 
\begin{eqnarray}
     P_f(-k) &=& \gamma^2 \tilde{P}_f(Rk) \gamma^2 \,, \nonumber \\
     P(k) &=& R P(Rk) R \,, \label{CPcond} \\
     \Gamma_\mu (-k'', -k') &=& - \gamma^2 R_{\mu\nu} \tilde{\Gamma}_\nu (Rk', Rk'') \gamma^2 \,, \nonumber
\end{eqnarray} 
then the value of each diagram equals the value of the CP-transformed diagram.

To obtain this result, we use relations (\ref{ePrel}) and
\begin{equation}
     u^{5 - \alpha}(-\vec{p}\,) = -\eta(\alpha) \gamma^2 \gamma^4 u^{\alpha*}(\vec{p}\,) \,,
     \label{CP1}
\end{equation}
and argue as in the case of C and P transformations.

\subsection{CT transformation}

CT transformation is the result of the combined application of C and T transformations, taken in any order. The CT-transformed graph is the mirror image of the original graph where we (a)~remirrored the lettering, (b)~replaced momenta $p_j$ with $Rp_j$ and $q_n$ with $Rq_n$, and (c)~replaced spin numbers $\alpha_j$ with $5 - \beta(\alpha_j)$ and $\ell_n$ with $-\ell_n$. If
\begin{eqnarray}
     P_f(k) &=& \gamma^4 \gamma^5 P_f(-Rk) \gamma^5\gamma^4 \,, \nonumber \\
     P(k) &=& R P(-Rk) R  \,, \label{CTcond} \\
     \Gamma_\mu(k', k'') &=& -\gamma^4 \gamma^5  R_{\mu\nu} \Gamma_\nu(-Rk', -Rk'') \gamma^5 \gamma^4 \,, \nonumber
\end{eqnarray} 
then the value of each diagram equals the value of the $CT$-transformed diagram multiplied by the phase factor
\begin{equation}
     (-1)^{S_+} \,,
     \label{CTpower}
\end{equation}
common to all diagrams of the same process.

To obtain this result, we use relations (\ref{produkt}), (\ref{TR2}), (\ref{TR1}), and
\begin{equation}
     u^{5 - \beta(\alpha)}(-\vec{p}\,) = (-1)^\alpha \eta(\alpha) \gamma^4 \gamma^5 u^{\alpha}(\vec{p}\,) \,, 
     \label{CT1}
\end{equation}
and argue by analogy to the case of P transformation.

\subsection{PT transformation}

The graph of the PT-transformed diagram is the result of the combined application of $P$ and $T$ transformations, taken in any order: It is the mirror image of the original graph where we (a)~remirrored the lettering, (b)~reversed the direction of all arrows on fermion lines, and (c)~replaced the spin numbers $\alpha_j$ with $\beta(\alpha_j)$ and $\ell_n$ with $-\ell_n$. If
\begin{eqnarray}
      P_f(k) &=& \gamma^2 \gamma^4 \gamma^5 \tilde{P}_f(k) \gamma^5 \gamma^4\gamma^2 \,, \nonumber \\
      P(k) &=& \tilde{P}(k) \,,      \label{PTcond} \\
     \Gamma_\mu(k', k'') &=& \gamma^2 \gamma^4 \gamma^5 \tilde{\Gamma}_\mu(k'', k') \gamma^5 \gamma^4 \gamma^2 \,, \nonumber
\end{eqnarray} 
then the value of each diagram equals the value of the $PT$-transformed diagram multiplied by the phase factor
\begin{equation}
     (-1)^{S_+} \,,
     \label{PTpower}
\end{equation}
common to all diagrams of the same process.

To check condition (\ref{PTcond}), we use relations (\ref{produkt}), (\ref{TR1}),
\begin{equation}
     u^{\beta(\alpha)}(\vec{p}\,) = (-1)^\alpha \gamma^2 \gamma^5 u^{\alpha*}(\vec{p}\,) \,, \qquad {\rm and} \qquad
     e^{-\ell}_\mu(\vec{q}) = e^{\ell *}_\mu(\vec{q}\,) \,, 
     \label{PT1}
\end{equation}
and argue as in the case of $C$ transformation.

\subsection{CPT transformation}

The graph of the CPT-transformed diagram is the result of the combined application of $C$, $P$, and $T$ transformations, taken in any order: It is the mirror image of the original graph where we remirrored the lettering, and replaced the spin numbers $\alpha_j$ with $5 - \beta(\alpha_j)$ and $\ell_n$ with $-\ell_n$. If
\begin{eqnarray}
     P_f(-k) &=& \gamma^5 P_f(k)\gamma^5 \,, \nonumber \\
     P(-k) &=& P(k) \,,  \label{CPTcond} \\
     \Gamma_\mu(-k',-k'') &=& - \gamma^5 \Gamma_\mu(k', k'') \gamma^5 \,, \nonumber
\end{eqnarray} 
then the value of each diagram equals the value of the CPT-transformed diagram multiplied by the phase factor 
\begin{equation}
     (-1)^{E_p + N_+ + S_+} \,,
     \label{CPTpower}
\end{equation}
common to all diagrams of the same process.

To check condition (\ref{CPTcond}), we use relations (\ref{produkt}), (\ref{TR1}), (\ref{PT1}), and
\begin{equation}
     u^{5 - \beta(\alpha)}(\vec{p}) = -(-1)^\alpha \gamma^5 u^\alpha(\vec{p}) \,,
     \label{CPT1}
\end{equation}
and argue by analogy to the case of $C$ transformation.

\section{Order-dependent invariances}
\label{secinvariances}

When transformed by the active, discrete transformations considered in the preceding Section, the value of each diagram has been invariant up to a phase factor uniquely determined by the external line factors, i.e. solely by the process under study. Now we will consider eight active, discrete transformations under which values of diagrams change into complex-conjugated values up to a phase factor determined both by the number of internal lines of the graph and by the process under study. These transformations are based on transformation H that is related to the Hermiticity of the fermion and photon propagators, and of the fermion-photon vertex function. Combining H with C, P, T, CP, CT, PT, and CPT transformations we obtain seven additional order-dependent active, discrete transformations of diagrams. To establish conditions under which each of these seven transformations is applicable, we combine the proofs of the properties of the H transformation and of the corresponding discrete transformation.

\subsection{Hermiticity-based transformation}

By definition, the \it Hermiticity-based \rm transformation H changes each incoming [outgoing] particle into an outgoing [incoming] particle with the same momentum and spin; and the H-transformed graph is the mirror image of the original graph where we remirrored the arrows and lettering. If the fermion and photon propagators and the vertex function are Hermitian in the sense that 
\begin{eqnarray}
     P_f^\dagger(k) &=& \gamma^4 P_f(k) \gamma^4 \,, \nonumber \\
     P(k) &=& P^\dagger(k) \,, \label{Hcond} \\
     \Gamma_\mu^\dagger( -k'', -k') &=& R_{\mu\nu} \gamma^4 \Gamma_\nu(k', k'') \gamma^4 \,, \nonumber
\end{eqnarray} 
then the value of each diagram equals the complex-conjugated value of the H-transformed diagram multiplied by the phase factor
\begin{equation}
     (-1)^{I} 
     \label{Hpower}
\end{equation}
that is not common to all diagrams of the same process, by (\ref{Imin}). To check conditions (\ref{Hcond}), we use equations (\ref{Pppro}), (\ref{Gvalue1})--(\ref{diag5}), and (\ref{CC5}).

Among the discrete transformations we considered so far, H is the first transformation that directly alters the numerical factor (\ref{Gvalue1}), by complex-conjugating it; which is the origin of the phase factor (\ref{Hpower}).

If the vertex function $\Gamma(k', k'')$ satisfies (\ref{Hcond}), then the interaction it describes is Hermitian in the sense that
\begin{displaymath}
     -(2\pi)^{-8}ie\int d_4k' d_4k'' d_4x_1 d_4x_2 d_4x_3 e^{ ik'( x_3 - x_1) + ik''( x_3 - x_2)}
            \bar{\psi}(x_1) \Gamma_\mu(-k', k'') A_\mu(x_3) \psi(x_2)
\end{displaymath}
is real for each bispinor field $\psi(x)$ and each real four-vector field $A_\mu(x)$; see \cite{Hooft}, Eqs. (2.1)--(2.6). This interaction is local in coordinate space if $\Gamma(k', k'')$ is a polynomial in $k'$ and $k''$.

Discrete transformations H, C, P, and T commute with each other; and applying each of them twice, we obtain the original process [diagram].

\subsection{HC transformation}

Transformation HC is the result of the combined applications of H and C transformations, taken in any order; an HC-transformed graph is the mirror image of the original graph where we remirrored the lettering and replaced spin numbers $\alpha_j$ with $5- \alpha_j$. If
\begin{eqnarray}
     P_f(k) &=& \gamma^2 P_f^*(-k) \gamma^2 \,, \nonumber \\
     P(k) &=& R P^*(-k) R \,, \label{HCcond} \\
     \Gamma_\mu(k', k'') &=& \gamma^2 R_{\mu \nu} \Gamma_\nu^* (k', k'') \,, \nonumber
\end{eqnarray} 
then the value of each HC-transformed diagram equals the complex-conjugated value of the original diagram multiplied by the phase factor
\begin{equation}
     (-1)^{I+ E_p + N_+} \,.
     \label{HCpower}
\end{equation}

\subsection{HP transformation}

An HP-transformed graph is the result of the combined applications of H and P transformations, taken in any order: It is the mirror image of the original graph where we remirrored the lettering and arrows of fermion lines, and replaced the momenta $p_j$ with $R p_j$ and $q_n$ with $R q_n$. If
\begin{eqnarray}
     P_f(k) &=& P_f^\dagger(R k) \,, \nonumber \\
     P(k) &=& P^\dagger(Rk) \,,  \label{HPcond} \\
     \Gamma_\mu(k', k'') &=& \Gamma_\mu^\dagger ( -R k'', -Rk') \,, \nonumber
\end{eqnarray} 
then the value of each HP-transformed diagram equals the complex-conjugated value of the original diagram multiplied by the phase factor
\begin{equation}
     (-1)^{I+ E_p + N_+} \,.
     \label{HPpower}
\end{equation}

\subsection{HT transformation}

An HT-transformed graph is the result of the combined applications of H and T transformations, taken in any order: It is obtained by replacing the momenta $p_j$ with $R p_j$ and $q_n$ with $R q_n$, and the spin numbers $\alpha_j$ with $\beta(\alpha_j)$ and $\ell_n$ with $-\ell_n$. If
\begin{eqnarray}
     P_f(k) &=& \gamma^2\gamma^4\gamma^5 P_f^*(R k) \gamma^5\gamma^4\gamma^2 \,, \nonumber \\
     P(k) &=& P^*(Rk) \,, \label{HTcond} \\
     \Gamma_\mu(k', k'') &=& \gamma^2\gamma^4\gamma^5 \Gamma_\mu^* ( -R k', -Rk'') \gamma^5\gamma^4\gamma^2 \,, \nonumber
\end{eqnarray} 
then the value of each HT-transformed diagram equals the complex-conjugated value of the original diagram multiplied by the phase factor
\begin{equation}
     (-1)^{I+ E_p + N_+ + S_+} \,.
     \label{HTpower}
\end{equation}

\subsection{HCP transformation}

An HCP-transformed graph is the result of the combined applications of H, C, and P transformations, taken in any order: It is the mirror image of the original graph where we remirrored the lettering, replaced the momenta $p_j$ with $R p_j$ and $q_n$ with $R q_n$, and replaced the spin numbers $\alpha_j$ with $5 - \alpha_j$. If
\begin{eqnarray}
     P_f(k) &=& \gamma^2\gamma^4 P_f^*(-R k) \gamma^4\gamma^2 \,, \nonumber \\
     P(k) &=& P^*(-Rk) \,, \label{HCPcond} \\
     \Gamma_\mu(k', k'') &=& -\gamma^2\gamma^4 \Gamma_\mu^* ( R k', Rk'') \gamma^4\gamma^2 \,, \nonumber
\end{eqnarray} 
then the value of each HCP-transformed diagram equals the complex-conjugated value of the original diagram multiplied by the phase factor
\begin{equation}
     (-1)^{I} \,.
     \label{HCPpower}
\end{equation}

\subsection{HCT transformation}

An HCT-transformed graph is the result of the combined applications of H, C, and T transformations, taken in any order: It is obtained from the original graph by reversing directions of arrows on fermion lines, replacing the momenta $p_j$ with $R p_j$ and $q_n$ with $R q_n$, and replacing the spin numbers $\alpha_j$ with $5 - \beta(\alpha_j)$ and $\ell_n$ with $-\ell_n$. If
\begin{eqnarray}
     P_f(k) &=& \gamma^5 P_f^\dagger(-R k) \gamma^5 \,, \nonumber \\
     P(k) &=& P^\dagger(-Rk) \,, \label{HCTcond} \\
     \Gamma_\mu(k', k'') &=& -\gamma^5 \Gamma_\mu^\dagger ( R k'', Rk') \gamma^5 \,, \nonumber
\end{eqnarray} 
then the value of each HCT-transformed diagram equals the complex-conjugated value of the original diagram multiplied by the phase factor
\begin{equation}
     (-1)^{I + S_+} \,.
     \label{HCTpower}
\end{equation}

\subsection{HPT transformation}

An HPT-transformed graph is the result of the combined applications of H, P, and T transformations, taken in any order: It is obtained from the original graph by replacing the spin numbers $\alpha_j$ with $\beta(\alpha_j)$ and $\ell_n$ with $-\ell_n$. If
\begin{eqnarray}
     P_f(k) &=& \gamma^2\gamma^5 P_f^*(k) \gamma^5\gamma^2 \,, \nonumber \\
     P(k) &=& R P^*(k) R \,, \label{HPTcond} \\
     \Gamma_\mu(k', k'') &=&  \gamma^2\gamma^5 R_{\mu\nu}\Gamma_\nu^* ( -k', -k'') \gamma^5 \gamma^2 \,, \nonumber
\end{eqnarray} 
then the value of each HPT-transformed diagram equals the complex-conjugated value of the original diagram multiplied by the phase factor
\begin{equation}
     (-1)^{I + S_+} \,.
     \label{HPTpower}
\end{equation}

\subsection{HCPT transformation}

An HCPT-transformed graph is the result of the combined applications of H, C, P, and T transformations, taken in any order: It is obtained by reversing directions of arrows on fermion lines, and replacing the spin numbers $\alpha_j$ with $5 - \beta(\alpha_j)$ and $\ell_n$ with $-\ell_n$. If
\begin{eqnarray}
     P_f(k) &=& \gamma^4\gamma^5 P_f^\dagger(-k) \gamma^5\gamma^4 \,, \nonumber \\
     P(k) &=& R P^\dagger(-k) R \,, \label{HCPTcond} \\
     \Gamma_\mu(k', k'') &=&  -\gamma^4\gamma^5 R_{\mu\nu}\Gamma_\nu^\dagger ( k'', k') \gamma^5 \gamma^4 \,, \nonumber
\end{eqnarray} 
then the value of each HCPT-transformed diagram equals the complex-conjuga\-ted value of the original diagram multiplied by the phase factor
\begin{equation}
     (-1)^{I + E_p + N_+ + S_+} \,.
     \label{HCPTpower}
\end{equation}

Table~\ref{prvatabla} gives actions on graphs of all 15 discrete transformations considered. Each of them conserves $E_f$, $E_p$, and the total energy of each process, and also $I_f$, $I_p$, $V$, $L_f$ and $L_i$ of each diagram.

\begin{table}
\centerline{
\vbox{\tabskip=0pt\offinterlineskip
\def\tablerule{\noalign{\hrule}}
\halign to300pt{\strut#&\vrule#\tabskip=1em plus2em&
   \hfil#& \vrule#& \hfil#\hfil& \vrule#& \hfil#\hfil& \vrule#& \hfil#\hfil& 
       \vrule#& \hfil#\hfil& \vrule#& \hfil#& \vrule#\tabskip=0pt\cr\tablerule
   &&\omit\hidewidth Symmetry\hidewidth&& \omit\hidewidth mirror\hidewidth&& \omit\hidewidth arrows\hidewidth&&
        \omit\hidewidth$\vec{p}$, $\vec{q}$ \hidewidth&& \omit\hidewidth $\alpha$\hidewidth&& 
        \omit\hidewidth $\ell$\hidewidth
   &\cr\tablerule
   &&C   &&   &&$*$&&   &&$5-\alpha       $&&$ \ell$&\cr\tablerule
   &&P   &&   &&   &&$*$&&$\alpha         $&&$ \ell$&\cr\tablerule
   &&T   &&$*$&&$*$&&$*$&&$\beta(\alpha)  $&&$-\ell$&\cr\tablerule
   &&CP  &&   &&$*$&&$*$&&$5-\alpha       $&&$ \ell$&\cr\tablerule
   &&CT  &&$*$&&   &&$*$&&$5-\beta(\alpha)$&&$-\ell$&\cr\tablerule
   &&PT  &&$*$&&$*$&&   &&$\beta(\alpha)  $&&$-\ell$&\cr\tablerule
   &&CPT &&$*$&&   &&   &&$5-\beta(\alpha)$&&$-\ell$&\cr\tablerule
   &&H   &&$*$&&$*$&&   &&$\alpha         $&&$ \ell$&\cr\tablerule
   &&HC  &&$*$&&   &&   &&$5-\alpha       $&&$ \ell$&\cr\tablerule
   &&HP  &&$*$&&$*$&&$*$&&$\alpha         $&&$ \ell$&\cr\tablerule
   &&HT  &&   &&   &&$*$&&$\beta(\alpha)  $&&$-\ell$&\cr\tablerule
   &&HCP &&$*$&&   &&$*$&&$5-\alpha       $&&$ \ell$&\cr\tablerule
   &&HCT &&   &&$*$&&$*$&&$5-\beta(\alpha)$&&$-\ell$&\cr\tablerule
   &&HPT &&   &&   &&   &&$\beta(\alpha)  $&&$-\ell$&\cr\tablerule
   &&HCPT&&   &&$*$&&   &&$5-\beta(\alpha)$&&$-\ell$&\cr\tablerule}}}
\caption[prva tabela]{For each of the 15 discrete transformations considered, columns 2-6 of the corresponding row describe its action on the graph. An asterisk $*$ in the second column signifies that it mirrors the graph and remirrors the lettering; an $*$ in the third column signifies that it reverses the direction of all arrows on fermion lines; an $*$ in the fourth column signifies that it reverses the direction of three-momentum of each particle; and in the fifth and sixth column, we write down the transformed values of the fermion and photon spin numbers $\alpha$ and $\ell$.}
\label{prvatabla}
\end{table}

\section{Lorentz transformation}
\label{secgeneral}

\subsection{Proper ortochronous Lorentz transformations of diagrams}

By definition, the value of a diagram transformed by a \it passive, proper ortochronous Lorentz transformation, \rm L, is given by (\ref{Gvalue1})--(\ref{diag5}) where we replaced (a)~the momenta $p_{j}$ and $q_{n}$ of external lines with $L p_{j}$ and $L q_{n}$ in factors (\ref{delta1}), and (b)~functions
\begin{equation}
     u^\alpha(\vec{p}) \quad {\rm with} \quad X(L) u^\alpha(\vec{p}) \,, \qquad
     e^\ell(\vec{q}) \quad {\rm with} \quad L e^\ell(\vec{q}) \,.
     \label{Lprop2}
\end{equation}
We will refer to this expression as the L-transformed value. If 
\begin{eqnarray}
     P_f(k) &=& X(L) P_f(L^{-1} k) X^{-1}(L) \,, \nonumber \\
     P(k)   &=& L P(L^{-1} k) L^{-1} \,, \label{Lpinv} \\
     \Gamma_\mu(k', k'') &=& L_{\mu\nu} X(L) \Gamma_\nu(L^{-1}k', L^{-1}k'') X^{-1}(L) \nonumber
\end{eqnarray}
for all $L$, then the value of each diagram equals the L-transformed value.

To prove that the value of each diagram is L-invariant if $P_f(k)$, $P(k)$ and $\Gamma(k', k'')$ satisfy (\ref{Lpinv}), we  start with the L-transformed value. 
\begin{itemize}
\item[(i)] In the changed factors (\ref{diag1})--(\ref{diag6}), we replace $P_f(k)$, $P(k)$ and $\Gamma(k', k'')$ with the right-hand sides of relations (\ref{Lpinv}), and take into account that $\gamma^4 X^\dagger(L) \gamma^4 = X^{-1}(L)$, ${\rm Tr}AB = {\rm Tr}BA$, $L$ is real, and $\tilde{L} = L^{-1}$. All factors $X(L)$ and $L_{\mu\nu}$, including those introduced by replacement (\ref{Lprop2}), cancel, and the resulting factors (\ref{diag1})--(\ref{diag6}) are of the original form (\ref{diag1})--(\ref{diag6}) with momenta $k_m$ replaced with $L^{-1}k_m$.
\item[(ii)] We replace all integrations over internal momenta $k_m$ with integrations over new variables $L^{-1} k_m$---which leaves four-integrals (\ref{facint}) unchanged---and then use relation $\delta_4\Bigl( L(\cdots) \Bigr) = \delta_4(\cdots)$.
\end{itemize}
So, the value of the L-transformed diagram equals the value of the original diagram.

\subsection{Lorentz form-invariant propagators}

We will establish that the most general fermion and photon propagators $P_f(k)$ and $P(k)$ that satisfy relations (\ref{Lpinv}) for each L can be written as
\begin{eqnarray}
     P_f(k) &=& f_1(k^2) + i\ksl f_2(k^2) + i\gamma^5 f_3(k^2) + \gamma^5 \ksl f_4(k^2) \,, \label{noro1} \\
     P_{\mu\nu}(k) &=& \delta_{\mu\nu} f_5(k^2) + k_\mu k_\nu f_6(k^2) \,, \label{noro6}
\end{eqnarray}
where $\ksl = \gamma^\mu k_\mu$, and $f_1(k^2)$, $f_2(k^2)$, $f_3(k^2)$, $f_4(k^2)$, $f_5(k^2)$, and $f_6(k^2)$ are arbitrary functions of $k^2$. 

So each Lorentz form-invariant photon propagator (\ref{noro6}) is symmetric and an even function of $k_\mu$. Propagators (\ref{noro1}) and (\ref{noro6}) are Hermitian, i.e. they satisfy relations (\ref{Hcond}), if, and only if
\begin{equation}
     f_i^* (k^2) = f_i (k^2) \qquad {\rm for} \quad i = 1, 2, 3, 4, 5, 6.
     \label{noro5}
\end{equation}

To derive representation (\ref{noro1}), we expand the $4 \times 4$ matrix $P_f(k)$ in terms of $16$ linearly independent matrices,
\[
     P_f(k) = c(k) + c_\mu(k) \gamma^\mu + c_{\mu\nu}(k) \sigma_{\mu\nu} + c'(k) \gamma^5 + c'_\mu(k) \gamma^5 \gamma^\mu \,,
\]
where $\sigma_{\mu\nu} = {1\over 4} (\gamma^\mu \gamma^\nu - \gamma^\nu \gamma^\mu )$ and $c_{\mu\nu}(k) = - c_{\nu\mu}(k)$, and conclude that $P_f(k)$ satisfies relation (\ref{Lpinv}) if, and only if,
\begin{eqnarray*}
     & c(k) = c(Lk) \,, \qquad c_\mu(k) = c_{\rho}(Lk) L_{\rho\mu} \,,  \\
     & c_{\mu\nu}(k) = c_{\rho\sigma}(Lk) L_{\rho\mu} L_{\sigma\nu} \,, \quad 
     c'(k) = c'(Lk) \,, \quad   c'_\mu(k) = c'_{\rho}(Lk) L_{\rho\mu} \,,     
\end{eqnarray*}
see \cite{Veltman}, App.~B and Sec.~4.2. Following Veltman \cite{Veltman}, Sec.~5.6, we now apply these relations to Lorentz transformation $S(k)$ that is the product of a space rotation $S$ and of the Lorentz transformation $L(k)$ defined by
\[
     L(k)q = \Bigl( \vec{q} + (\gamma - 1) \hat{k} (\hat{k}\cdot \vec{q}) -
          \sqrt{ \gamma^2 - 1} \hat{k} q_0 ,\, - i\sqrt{ \gamma^2 - 
          1}\hat{k} \cdot \vec{q} + i\gamma q_0 \Bigr) \,,
\]
where $k_0^2 > \vec{k}^2$, $\gamma = k_0/\sqrt{-k^2}$, and $\hat{k} = \vec{k}/|\vec{k}|$; $S(k)$ is proper and ortochronous for $k_0 > 0$, and such that $S(k) k = k^{(0)} = (0, 0, 0, i\sqrt{-k^2} )$. We obtain relations
\begin{eqnarray*}
     & c(k) = c(k^{(0)}) \,, \quad  c_\mu(k) = c_{\rho}(k^{(0)}) S_{\rho\mu}(k) \,,  \\
     & c_{\mu\nu}(k) = c_{\rho\sigma}(k^{(0)}) S_{\rho\mu}(k) S_{\sigma\nu}(k) \,, \quad 
     c'(k) = c'(k^{(0)}) \,, \quad  c'_\mu(k) = c'_{\rho}(k^{(0)}) S_{\rho\mu}(k) \nonumber
\end{eqnarray*}
valid for each space rotation $S$. They imply that
\[
     c_{ij}(k^{(0)}) = 
     c_i(k^{(0)}) =  c_{4j}(k^{(0)}) = c_{i4}(k^{(0)}) = c'_i(k^{(0)}) = 0 \,, 
\]
$i, j = 1, 2, 3$, where we used the antisymmetry of $c_{\mu\nu}(k^{(0)})$; so, $P_f(k)$ may be written as (\ref{noro1}). 

To derive (\ref{noro6}) from (\ref{Lpinv}), we proceed similarly and conclude that 
\[
     P_{ij}(k^{(0)}) = P_{11}(k^{(0)}) \delta_{ij}\,, \qquad
     P_{i4}(k^{(0)}) = P_{4j}(k^{(0)}) = 0 \,, \quad i, j = 1, 2, 3 \,, 
\]
which implies (\ref{noro6}).

\subsection{Lorentz form-invariant vertex function}

One can represent the most general fermion-photon vertex function  $\Gamma_\mu(k',k'')$ that satisfies relation (\ref{Lpinv}) in terms of 24 basis tensors as given by Bernstein\cite{Bernstein}. Ball and Chiu\cite{Ball} used the first 12 of them to construct the basis convenient to represent a general form of fermion-photon vertex consistent with the Ward-Takahashi identities and free of kinematic singularities. Kizilers\"u et al.\cite{Kizilersu} modified this basis. We found both bases equally convenient, and chose the latter one to represent the most general Lorentz form-invariant fermion-photon vertex function $\Gamma_\mu(k',k'')$ as
\begin{equation}
     \Gamma_\mu(k',k'') = \sum_{i=1}^{24} F_i(k^2, q^2, k\cdot q) \Gamma_\mu^i(k,q) 
     \label{nor1}
\end{equation}
in terms of 24 arbitrary complex-valued scalar functions $F_i(k^2, q^2, k\cdot q)$, $k = k' + k''$, $q = k'' - k'$, and 24 basis tensors:
\begin{eqnarray}
     \Gamma_\mu^1(k,q) &=& \gamma^\mu \,, \cr
     \Gamma_\mu^2(k,q) &=& k_\mu \,, \cr
     \Gamma_\mu^3(k,q) &=& k_\mu \ksl \,, \cr
     \Gamma_\mu^4(k,q) &=& q_\mu \sigma_{\rho\tau} q_\rho k_\tau \,, \cr
     \Gamma_\mu^5(k,q) &=& {\textstyle{1\over 2}}[ q^2 k_\mu - (k \cdot q) q_\mu ] \,, \cr
     \Gamma_\mu^6(k,q) &=& \Gamma_\mu^5(k, q) \ksl \,, \cr
     \Gamma_\mu^7(k,q) &=& q^2 \gamma^\mu - q_\mu \qsl \,, & \label{nor2}\cr
     \Gamma_\mu^8(k,q) &=& 2( q^2 \sigma_{\mu\nu} k_\nu +  q_\mu \sigma_{\rho\tau} k_\rho q_\tau )\,, \cr
     \Gamma_\mu^9(k,q) &=& 2\sigma_{\nu\mu} q_\nu \,, \cr
     \Gamma_\mu^{10}(k,q) &=& k_\mu \qsl - (k\cdot q) \gamma^\mu \,, \cr
     \Gamma_\mu^{11}(k,q) &=& k_\mu \sigma_{\rho\tau} q_\rho k_\tau - (k\cdot q) \sigma_{\mu\nu} k_\nu \,, \cr
     \Gamma_\mu^{12}(k,q) &=& {\textstyle{1\over 2}}(\gamma^\mu \sigma_{\rho\tau} + 
               \sigma_{\rho\tau}\gamma^\mu) k_\rho q_\tau \,, \cr
     \Gamma_\mu^i(k,q) &=& \gamma^5 \Gamma_\mu^{i-12}(k,q) \,, \quad i = 13, 14, \ldots, 24 \,.
\end{eqnarray}

The vertex function $\Gamma(k', k'')$ satisfies Hermiticity condition (\ref{Hcond}) if, and only if 
\begin{eqnarray}
     F_i(k^2, q^2, -k\cdot q) &=& F_i^*(k^2, q^2, k\cdot q) \,, \quad i \in I_1 \cup I_2 \,, \cr
     F_i(k^2, q^2, -k\cdot q) &=& - F_i^*(k^2, q^2, k\cdot q) \,, \quad i \in I_3 \cup I_4  \,,
     \label{verHerres}
\end{eqnarray}
where
\begin{eqnarray}
    I_1 = \{1,2,3,5,7,9,11,12 \} \,, \qquad && I_2 = \{13,15,16,18,19,20,24 \} \nonumber \\
    I_3 = \{4,8,10 \} \,,\qquad && I_4 = \{14,17,21,22,23 \} \,. \label{Indef}
\end{eqnarray}

\subsection{Invariances of diagrams}

We now apply conditions of Section~\ref{secsymmetries} to the Lorentz form-invariant propagators (\ref{noro1}) and (\ref{noro6}) and vertex function (\ref{nor1}) to obtain conditions on functions $f_i(k^2)$ and $F_i(k^2, q^2, k\cdot q)$ under which L-invariant diagrams are invariant under C, P, T, CP, CT, PT, and CPT transformations. 

\it Charge conjugation. \rm The propagators (\ref{noro1}) and (\ref{noro6}) and the vertex function (\ref{nor1}) satisfy conditions (\ref{Ccond}) for invariance of diagrams under charge conjugation if, and only if
\begin{eqnarray}
     f_4(k^2) &=& 0 \,; \cr
     F_i(k^2, q^2, -k\cdot q) &=& F_i(k^2, q^2, k\cdot q) \,, \quad i \in I_1 \cup I_4 \,; \cr
     F_i(k^2, q^2, -k\cdot q) &=& -F_i(k^2, q^2, k\cdot q) \,,\quad i \in I_2 \cup I_3 \,. 
\end{eqnarray}

\it Space reflection. \rm The propagators (\ref{noro1}) and (\ref{noro6}) and the vertex function (\ref{nor1}) satisfy conditions (\ref{Pcond}) for invariance of diagrams under space reflection if, and only if
\begin{eqnarray}
     f_i(k^2) &=& 0 \,, \quad i= 3, 4; \cr
     F_i(k^2, q^2, k\cdot q) &=& 0 \,, \quad i \in I_2 \cup I_4 \,. 
\end{eqnarray}

\it Time reversal. \rm The propagators (\ref{noro1}) and (\ref{noro6}) and the vertex function (\ref{nor1}) satisfy conditions (\ref{Tcond}) for invariance of diagrams under time reversal if, and only if
\begin{eqnarray}
     f_3(k^2) &=& 0 \,; \cr
     F_i(k^2, q^2, -k\cdot q) &=& F_i(k^2, q^2, k\cdot q) \,, \quad i \in I_1 \cup I_2 \,; \cr
     F_i(k^2, q^2, -k\cdot q) &=& -F_i(k^2, q^2, k\cdot q) \,, \quad i \in I_3 \cup I_4  \,. 
\end{eqnarray}

\it CPT transformation. \rm The Lorentz form-invariant propagators (\ref{noro1}) and (\ref{noro6}) and vertex function (\ref{nor1}) always satisfy conditions (\ref{CPTcond}) for invariance of diagrams under CPT transformation. So each diagram with Lorentz form-invariant propagators and vertex function is also CPT-invariant; which is a special case of the \it CPT theorem. \rm As a consequence, if the Lorentz form-invariant propagators (\ref{noro1}) and (\ref{noro6}) and the vertex function (\ref{nor1}) satisfy conditions (\ref{Ccond}), (\ref{Pcond}), or (\ref{Tcond}) for invariance of diagrams under C, P, or T transformation, they satisfy also conditions (\ref{PTcond}), (\ref{CTcond}), or (\ref{CPcond}) for invariance under PT, CT, or CP transformation and vice versa, since applying twice C, P, or T transformation, we obtain the original diagram.

\it Hermiticity and invariance under L, C, P, and T transformations. \rm The QED propagators [with $i\epsilon = 0$] and vertex function satisfy conditions (\ref{Lpinv}), (\ref{Ccond}), (\ref{Pcond}), and (\ref{Tcond}) [(\ref{Hcond})] for invariance of Feynman diagrams under L, C, P, and T transformations [Hermiticity]. The regularizing propagators $P_f(k)$ and $P(k)$ and vertex function $\Gamma(k', k'')$ satisfy these conditions if, and only if they are of the form (\ref{noro1}), (\ref{noro6}) and (\ref{nor1}), and
\begin{eqnarray}
     f_i^*(k^2) &=& f_i(k^2) \,,  \quad i = 1, 2; \cr
     f_i(k^2) &=& 0 \,, \quad i = 3, 4, \cr
     F_i(k^2, q^2, -k\cdot q) &=& F_i(k^2, q^2, k\cdot q) \,, \quad i \in I_1 \,; \cr
     F_i(k^2, q^2, -k\cdot q) &=& -F_i(k^2, q^2, k\cdot q) \,, \quad i \in I_3 \,; \cr 
     F_i^*(k^2, q^2, k\cdot q) &=& F_i(k^2, q^2, k\cdot q) \,, \quad i \in I_1 \cup I_3 \,; \cr
     F_i(k^2, q^2, k\cdot q) &=& 0 \,, \quad i \in I_2 \cup I_4 \,.
\end{eqnarray}
The one-loop QED vertex calculated by Kizilers\"u et al.\cite{Kizilersu} is such.

\section{Properties of perturbative transition probabilities}
\label{secproperties}

As the value of each diagram is proportional to $e^V$, for each process we can calculate the transition probability up to the $n$-th order of $e^2$ (transition probability for short), $n \ge V_{{\rm min}}$, by taking the squared absolute value of the sum over all diagrams of this process that have up to $2n - V_{{\rm min}}$ vertices, and then omitting all higher-order terms; we obtain an expansion in powers of $e^2$, by (\ref{Vmin}). So for each process, the transition probability is invariant under the replacement of $e$ with $-e$.

\it Pauli exclusion principle. \rm Processes described by diagrams conform to the Pauli exclusion principle: For each process where two of incoming or two of outgoing fermions are the same, the transition probability vanishes to all orders, since then each diagram in the sum needed to calculate the transition probability appears twice but with opposite signs. So the transition probability for electron-electron scattering vanishes to all orders if two incoming or two outgoing electrons have the same momenta and spins.

\it Conservation of electric charge and of momentum. \rm According to Section~\ref{secconservation}, transition probability (a)~cannot be defined by diagrams for processes where the total electric charge is not conserved, e.g. for processes where the difference between the total numbers of incoming and outgoing fermions is odd; and (b)~vanishes for processes that do not conserve the total momentum, e.g. for processes involving two fermions and one photon since $\kappa \ne 2m$. So each non-vanishing diagram has at least two vertices.

\it Lorentz invariance. \rm If the fermion and photon propagators $P_f(k)$ and $P(k)$ and the fermion-photon vertex function $\Gamma(k', k'')$ satisfy relations (\ref{Lpinv}), then for each process the transition probability equals to all orders the transition probability for the process identical to the original process but observed from the L-transformed inertial reference frame. L-invariance implies that the total angular momentum is conserved, see e.g. \cite{Weinberg}, Sections 3.3 and 3.7. As a consequence, the transition probability vanishes for two-photon annihilation of an electron-positron pair at rest when $S_+ = 2$; Sakurai\cite{Sakurai} considered this case in detail.

\subsection{Symmetries}

For each discrete transformation X = C, P, T, CP, CT, PT, or CPT, we will regard two processes as X-related if each X-transformed diagram of one process is a diagram of the other process; and we will call a process X-invariant if the X-related process is the same as the original process. Using results of Section~\ref{secsymmetries}, we will give conditions under which transition probabilities of two X-related processes are equal. To this end we note: If we take the sum of diagrams we need to calculate the transition probability for some process and X-transform each diagram, we obtain the sum of diagrams we need for calculating the transition probability for the X-related process, since each transformation conserves the number of vertices and does nothing if applied twice. So these two sums are identical for each X-invariant process.  

If a sum of diagrams of a process is invariant under X-transformation, it vanishes if the corresponding phase factor equals $-1$, i.e. if it's exponent is odd. In this connection we note that for each integer $m$ the sums $m + N_+$ and $m + N_-$ are both either odd or even, by (\ref{nspm}); and the same goes for $m + S_+$ and $m + S_-$. So in this respect, $N_+$ and $N_-$, or $S_+$ and $S_-$, are two equivalent characteristics of the process.

We will now state characteristic properties of the seven discrete transformations and collate them in Table~\ref{drugatabla}.

\it Charge conjugation symmetry. \rm For each process, the C-related process is obtained by replacing each particle by its antiparticle with the same momentum and spin. Transition probabilities for two C-related processes are equal to all orders if the fermion propagator $P_f(k)$ and the fermion-photon vertex function $\Gamma(k', k'')$ satisfy relations (\ref{Ccond}).

If condition (\ref{Ccond}) is satisfied, the transition probability vanishes to all orders for each C-invariant process if $E_p + N_+$ is odd. In particular, when condition (\ref{Ccond}) is satisfied: (A)~The transition probability for a two-photon annihilation of an electron-positron pair vanishes to all orders for each process where the electron and the positron have the same momenta and spins \cite{Sakurai}. (B)~There is \it Furry's theorem \rm \cite{Weinberg}, Sects.~3.3 and 10.1: In computing a transition probability we may omit each diagram with a subdiagram that has an odd number of external photon lines and no external fermion lines; e.g. if condition (\ref{Ccond}) is satisfied, we may omit all diagrams that contain a tadpole, or a closed fermion loop with an odd number of vertices; for an example see Fig.~\ref{graf}. (C)~Transition probability for any process with an odd number of photons vanishes to all orders in the absence of fermions; i.e. in the absence of fermions the transition probability vanishes if the products of the photon charge-conjugation parities in the initial and final states are not equal. Which is also a consequence of Furry's theorem.

As a consequence of the conservation of momentum and charge, and of Furry's theorem, for each process involving only three particles the transition probability either cannot be defined by diagrams, or it vanishes to all orders. And the same goes for each process that involves only an odd number of photons or fermions; cf.~\cite{Weinberg}, Sect.~3.3.

\it Space reflection symmetry. \rm For each process, the P-related process is obtained by reversing the direction of three-momentum of each particle. Transition probability for two P-related processes are equal to all orders if the fermion and photon propagators $P_f(k)$ and $P(k)$ and the fermion-photon vertex function $\Gamma(k', k'')$ satisfy relations (\ref{Pcond}). In such a case, the transition probability vanishes to all orders for each P-invariant process if $E_p + N_+$ is odd. So the transition probability for two-photon annihilation of an electron-positron pair vanishes if they are at rest and the photons have opposite three-momenta and the same spins\cite{Sakurai}.

\it Time reversal symmetry. \rm For each process, the T-related process is obtained by changing incoming [outgoing] particles into outgoing [incoming] ones, reversing the direction of their three-momenta, and inverting their spins. If the propagators $P_f(k)$ and $P(k)$ and the vertex function $\Gamma(k', k'')$ satisfy relations (\ref{Tcond}), then transition probability for two T-related processes are equal to all orders. In such a case, transition probability vanishes to all orders for each T-invariant process if $E_p + N_+ + S_+ $ is odd; an example is Compton scattering if the sum of three-momenta of incoming electron and photon vanishes and the scattering only inverts their spins and reverses their three-momenta.

\it CP symmetry. \rm For each process, the CP-related process is obtained by changing particles into antiparticles with the reversed direction of three-momen\-tum. If the propagators $P_f(k)$ and $P(k)$ and the vertex function $\Gamma(k', k'')$ satisfy relations (\ref{CPcond}), then for each process the transition probability equals to all orders the transition probability for the $CP$-related process.

\it CT symmetry. \rm For each process, the CT-related process is obtained by changing incoming [outgoing] particles into outgoing [incoming] antiparticles, reversing the direction of their three-momentum, and inverting their spin. If the propagators $P_f(k)$ and $P(k)$ and the vertex function $\Gamma(k', k'')$ satisfy relations (\ref{CTcond}), then for each process the transition probability equals to all orders the transition probability for the $CT$-related process. In such a case, the transition probability vanishes to all orders for each CT-invariant process if $S_+$ is odd; however, there is no such process.

\it PT symmetry. \rm For each process, the PT-related process is obtained by changing incoming [outgoing] particles into outgoing [incoming] ones, and inverting their spin. If the propagators $P_f(k)$ and $P(k)$ and the vertex function $\Gamma(k', k'')$ satisfy relations (\ref{PTcond}), then for each process the transition probability equals to all orders the transition probability for the $PT$-related process. In such a case, the transition probability vanishes to all orders for each PT-invariant process if $S_+$ is odd. So the transition probability for the Compton photon-electron scattering vanishes to all orders if the momenta of electron and photon remain the same and only spins are inverted.

\it CPT symmetry. \rm For each process, the CPT-related process is obtained by changing incoming [outgoing] particles into outgoing [incoming] antiparticles with inverted spin. If the propagators $P_f(k)$ and $P(k)$ and the vertex function $\Gamma(k', k'')$ satisfy relations (\ref{CPTcond}), then for each process the transition probability equals to all orders the transition probability for the $CPT$-related process. In such a case, the transition probability vanishes to all orders for each CPT-invariant process if $E_p + N_+ + S_+$ is odd; however, there is no such process.

\begin{table}[t]
\centerline{
\vbox{\tabskip=0pt \offinterlineskip
\def\tablerule{\noalign{\hrule}}
\halign to350pt{\strut#&\vrule#\tabskip=1em plus2em&
   \hfil#& \vrule#& \hfil#\hfil& \vrule#& \hfil#\hfil& \vrule#& \hfil#\hfil& \vrule#& \hfil#\hfil& 
       \vrule#& \hfil#\hfil& \vrule#& \hfil#& \vrule#\tabskip=0pt\cr\tablerule
   &&\omit\hidewidth Symmetry\hidewidth&& \omit\hidewidth in-out\hidewidth&& \omit\hidewidth $\vec{p}$, 
   $\vec{q}$ \hidewidth&& \omit\hidewidth spin\hidewidth&& \omit\hidewidth charge\hidewidth&& 
      \omit\hidewidth$(-1)^{\ldots}$\hidewidth&& \omit\hidewidth process\hidewidth
   &\cr\tablerule
   &&C&& && && &&$*$&&$E_p + N_+$&&$2$&\cr\tablerule
   &&P&& &&$*$&& && &&$E_p + N_+$&&$2$&\cr\tablerule
   &&T&&$*$&&$*$&&$*$&& &&$E_p + N_+ + S_+$&&$1$&\cr\tablerule
   &&CP&& &&$*$&& &&$*$&&$0$&&$0$&\cr\tablerule
   &&CT&&$*$&&$*$&&$*$&&$*$&&$S_+$&&$0$&\cr\tablerule
   &&PT&&$*$&& &&$*$&& &&$S_+$&&$1$&\cr\tablerule
   &&CPT&&$*$&& &&$*$&&$*$&&$E_p + N_+ + S_+$&&$0$&\cr\tablerule
   &&H&&$*$&& && && &&$I$&&$0$&\cr\tablerule
   &&HC&&$*$&& && &&$*$&&$E_p + N_+ + I$&&$0$&\cr\tablerule
   &&HP&&$*$&&$*$&& && &&$E_p + N_+ + I$&&$0$&\cr\tablerule
   &&HT&& &&$*$&&$*$&& &&$E_p + N_+ + S_+ + I$&&$0$&\cr\tablerule
   &&HCP&&$*$&&$*$&& &&$*$&&$I$&&$0$&\cr\tablerule
   &&HCT&& &&$*$&&$*$&&$*$&&$S_+ + I$&&$0$&\cr\tablerule
   &&HPT&& && &&$*$&& &&$S_+ + I$&&$0$&\cr\tablerule
   &&HCPT&& && &&$*$&&$*$&&$E_p + N_+ + S_+ + I$&&$0$&\cr\tablerule}}}
\caption[druga tabela]{For each of the 15 discrete transformations considered, columns 2-5 of the corresponding row describe how we have to transform the particles of a process to obtain the particles of the related process. An asterisk $*$ in the second column signifies that we have to change the original incoming particles into outgoing ones, and vice versa; an $*$ in the third column signifies that we have to reverse the direction of three-momentum of each particle; an $*$ in the fourth column signifies that we have to invert the spin of each particle (if photons are linearly polarized, then the polarization numbers of the original photons and of the photons of the transformed process are the same); and an $*$ in the fifth column signifies that we have to replace each particle with its antiparticle. Symbols in the sixth column give the exponent in the corresponding phase factor. A number $0$ in the seventh column signifies that there is no process where transition probability vanishes solely due to the corresponding symmetry. Whereas a number $1$ [$2$] indicates that there are particular values of momenta and spin numbers such that transition probability for Compton scattering [two-photon annihilation of positron-electron pair] vanishes due to the corresponding symmetry.}
\label{drugatabla}
\end{table}

\subsection{Linearly polarized photons}

Each incoming or outgoing linearly polarized massive photon is specified by its momentum $q$ and linear polarization number $1$, $2$, or $3$; the corresponding external-line factors are
\begin{eqnarray}
    &&e^{(1)}(\vec{q}\,) = \Bigl( e^1(\vec{q}\,) + e^{-1}(\vec{q}\,) \Bigr)/\sqrt2 \,,\nonumber \\
    &&e^{(2)}(\vec{q}\,) = i\Bigl( e^{-1}(\vec{q}\,) - e^1(\vec{q}\,) \Bigr)/\sqrt2 \,, \label{linee} \\
    &&e^{(3)}(\vec{q}\,) = e^0(\vec{q}\,) \,, \nonumber 
\end{eqnarray}
they are real, and so the same for incoming and outgoing photons, see\cite{Veltman}, Sec.~4.4.

For linearly polarized photons, the transformed graphs are by definition the same as for circularly polarized photons except that the linear polarization numbers are never changed. If the fermion and photon propagators $P_f(k)$ and $P(k)$ and the fermion-photon vertex function $\Gamma(k',k'')$ satisfy relations (\ref{Ccond}), (\ref{Pcond}), (\ref{Tcond}), (\ref{CPcond}), (\ref{CTcond}), (\ref{PTcond}), or (\ref{CPTcond}), then the value of each diagram equals the value of the transformed diagram multiplied by the same phase factor as in the case of circulary polarized photons.

So for each process involving linearly polarized photons the related process is obtained in the same way as the related process for circularly polarized photons except that polarization numbers of linearly polarized photons are not changed. As a consequence, there are results about properties of transition probabilities analogous to the preceding ones of this section. In particular, if the conditions (\ref{CPTcond}) for CPT symmetry are satisfied, then for each process without fermions the transition probability stays the same if we exchange the incoming and outgoing, linearly polarized photons; and it vanishes if their total number $E_p$ is odd.

\subsection{Symmetries of the lowest-order transition probabilities}

The QED fermion and photon propagators with $i\epsilon = 0$, and vertex $-ie\gamma^\mu$ are Hermitian, i.e. they satisfy relations (\ref{CHcon}). As they also satisfy relations (\ref{Ccond}), (\ref{Pcond}), and (\ref{Tcond}) for invariance under C, P, and T transformations, they satisfy all relations of Section~\ref{secinvariances} for invariance under H, HC, HP, HT, HCP, HCT, HPT, and HCPT transformations provided $i\epsilon = 0$. If $E_f \ne 0$, then in the lowest-order, tree diagrams there is no integration over independent loops by (\ref{Lmin}); and we have diagrams with $i\epsilon = 0$. So all QED tree transition probabilities are invariant under H, HC, HP, HT, HCP, HCT, HPT, and HCPT transformations.

\it Hermiticity-linked symmetry. \rm For each process, the H-related process is obtained by changing incoming [outgoing] particles into identical, outgoing [incoming] ones. If the fermion and photon propagators $P_f(k)$ and $P(k)$ and the fermion-photon vertex function $\Gamma(k', k'')$ satisfy relations (\ref{Hcond}), then the lowest-order transition probabilities for two H-related processes are equal.

\it HC symmetry. \rm For each process, the H-related process is obtained by changing each incoming [outgoing] particle into an outgoing [incoming] antiparticle with the same momentum and spin. If $P_f(k)$, $P(k)$, and $\Gamma(k', k'')$ satisfy relations (\ref{HCcond}), then the lowest-order transition probabilities for two HC-related processes are equal.

\it HP symmetry. \rm For each process, the HP-related process is obtained by changing each incoming [outgoing] particle into an outgoing [incoming] particle with the same spin and opposite three-momentum. If conditions (\ref{HPcond}) are satisfied, then the lowest-order transition probabilities for two HP-related processes are equal.

\it HT symmetry. \rm For each process, the HT-related process is obtained by reversing the direction of three-momentum of each particle and inverting its spin. If conditions (\ref{HTcond}) are satisfied, then the lowest-order transition probabilities for two HT-related processes are equal.

\it HCP symmetry. \rm For each process, the HCP-related process is obtained by changing each incoming [outgoing] particle into an outgoing [incoming] antiparticle with reversed direction of three-momentum and the same spin. If conditions (\ref{HCPcond}) are satisfied, then the lowest-order transition probabilities for two HCP-related processes are equal.

\it HCT symmetry. \rm For each process, the HCT-related process is obtained by changing each particle into its antiparticle, reversing the direction of its three-momentum, and inverting its spin. If conditions (\ref{HCTcond}) are satisfied, then the lowest-order transition probabilities for two HCT-related processes are equal.

\it HPT symmetry. \rm For each process, the HPT-related process is obtained by inverting the spin of each particle. If conditions (\ref{HPTcond}) are satisfied, then the lowest-order transition probabilities for two HPT-related processes are equal.

Symmetries analogous to HT and HPT were derived from unitarity of S-matrix by Weinberg\cite{Weinberg}, p.130, in connection with Watson's theorem.

\it HCPT symmetry. \rm For each process, the HCPT-related process is obtained by changing each particle into its antiparticle and inverting its spin. If conditions (\ref{HCPTcond}) are satisfied, then the lowest-order transition probabilities for two HCPT-related processes are equal.

Results (\ref{Hpower}), (\ref{HCpower}), (\ref{HPpower}), (\ref{HTpower}), (\ref{HCPpower}), (\ref{HCTpower}), (\ref{HPTpower}), and (\ref{HCPTpower}) about the phase factors imply the following result about the lowest-order transition amplitude: If $P_f(k)$, $P(k)$ and $\Gamma(k',k'')$ satisfy relations (\ref{Hcond}), (\ref{HCcond}), (\ref{HPcond}), (\ref{HTcond}), (\ref{HCPcond}), (\ref{HCTcond}), (\ref{HPTcond}), or (\ref{HCPTcond}), and the process is H, HC, HP, HT, HCP, HCT, HPT, or HCPT invariant, then the real or imaginary part of the lowest-order transition amplitude vanishes depending on whether the corresponding phase factor equals $-1$ or $1$.

Table~2 gives properties and examples of the fifteen discrete symmetries of transition probabilities. These fifteeen discrete symmetries cover all possible combinations of the following four transformations of fermions and photons: incoming $\leftrightarrow$ outgoing; $\vec{p}, \vec{q} \rightarrow -\vec{p}, -\vec{q}$; spin up $\leftrightarrow$ spin down; particle $\leftrightarrow$ antiparticle.

\section{Formal invariances}
\label{secformal}

We now consider invariances under three kinds of transformations such that the transformed diagrams are not diagrams of a physical process as specified in Sections 1 and \ref{secdiagrams}.

\subsection{Generalization of Lorentz transformation}

Arguing by analogy to the case of L-transformation, we can show that transition probability is invariant to all orders under the following substitutions:
\begin{eqnarray}
    u^{\alpha_j}(\vec{p}_j) &\to& e^{i\phi_j} M(p_j) u^{\alpha_j}(\vec{p}_j) \,, \qquad p_j \to O p_j \,, \nonumber \\
    e^{\ell_n}_\mu(\vec{q}_n) &\to& e^{i\phi_n} N(q_n) e^{\ell_n}_\mu(\vec{q}_n) \,, \qquad q_n \to O q_n \,, \nonumber \\
    P_f(k) &\to & e^{i\phi_f} (\gamma^4 M^\dagger(k) \gamma^4)^{-1} P_f(O k) M^{-1}(k) \,, \label{formtransf} \\
    P(k) &\to & e^{i\phi_p} N(k) P(O k) N^{-1}(k) \,, \nonumber \\
    \Gamma_\mu(k',k'') &\to& e^{i\phi_v} N_{\mu\nu}(k' - k'') M(k') \Gamma_\nu(Ok', Ok'') 
        \gamma^4M^\dagger(k'')\gamma^4 \,, \nonumber
\end{eqnarray}
where $M(k)$, $N(k)$, and $O$ are $4\times 4$ matrices such that $N^{-1} = N^\dagger$ and ${\rm det} O = 1$, and $\phi_j$, $\phi_n$, $\phi_f$, $\phi_p$, and $\phi_v$ are real constants such that $2\phi_j + \phi_p + 2\phi_v = 0$.

\subsection{Chiral transformation}

If
\begin{equation}
     P_f(k) = - \gamma^5 P_f(k) \gamma^5 \qquad {\rm and} \qquad
     \Gamma_\mu(k', k'') = - \gamma^5 \Gamma_\mu (k', k'') \gamma^5 \,,
     \label{CHcon}
\end{equation}
then the value of each diagram is invariant under \it chiral transformation, \rm i.e. under the replacement of $u^\alpha(\vec{p}\,)$ with $\gamma^5 u^\alpha(\vec{p}\,)$ in external fermion line factors (\ref{diag1}). A diagram transformed by the chiral transformation does not correspond to a physical process, by (\ref{CPT1}).

The general Lorentz-invariant propagators (\ref{noro1}) and (\ref{noro6}) and the vertex function (\ref{nor1}) satisfy conditions (\ref{CHcon}) for invariance of diagrams under chiral transformation if, and only if
\begin{eqnarray}
     f_i(k^2) &=& 0 \,, \quad i = 1, 3; \cr
     F_i(k^2, q^2, k\cdot q) &=& 0 \,, \quad i= 2, 4, 5, 8, 9, 11, 14, 16, 17, 20, 21, 23.
\end{eqnarray}

In contrast to conditions (\ref{Ccond}), (\ref{Pcond}), (\ref{Tcond}), (\ref{CPcond}) (\ref{CTcond}), (\ref{PTcond}), and (\ref{CPTcond}) for invariance under C, P, T, CP, CT, PT, and CPT transformations respectively that are satisfied by the QED fermion and photon propagators and vertex, condition (\ref{CHcon}) for invariance under chiral transformation is satisfied by the QED fermion propagator only asymptotically for very large momentum $k$ relative to the mass $m$.

\subsection{Formal Hermitian-based invariances}

If we replace $e$ with $\sqrt{-i}\,e$ in factor (\ref{Gvalue1}) for each diagram so that it equals
\begin{equation}
     (-1)^{L_f} i^{(E_f + E_p)/2} e^V (2\pi)^{4(V-I)} \,,
\end{equation}
and if the propagators $P_f(k)$ and $P(k)$ and the vertex function $\Gamma(k', k'')$ satisfy relations (\ref{Hcond}), (\ref{HCcond}), (\ref{HPcond}), (\ref{HTcond}), (\ref{HCPcond}), (\ref{HCTcond}), (\ref{HPTcond}), and (\ref{HCPTcond}), then for each process the transition probability equals to all orders the transition probability for the H-, HC-, HP-, HT-, HCP-, HCT-, HPT-, and HCPT-related process respectively.

\section{Concluding remarks}
\label{secconclusions}

Together with many physicists we share the idea that Feynman diagrams for a scattering process contain more truth about this process than the underlying formalism used to derive them and are, therefore, likely to outlive it, cf. e.g. \cite{Hooft,Veltman,Bjorken}. If so, in the spirit of 't Hooft and Veltman \cite{Hooft}, diagrams may be taken as the basis for a bottom-up approach to potential theories that underly the standard model and do not require regularization. So far mainly those properties of regularized Feynman diagrams were of interest that optimize the renormalization calculations. However, if we start looking for regularized Feynman diagrams such that all parameters can be regarded as physical constants, then a much wider class of properties is of interest. Which motivated this paper.

We studied invariances of connected QED diagrams that were made finite (i.e. regularized) by the replacement of the QED propagators and vertex factor with regularizing ones. We gave properties of regularizing propagators and vertex function that are sufficient for the corresponding transition amplitudes to have the same symmetries as the original QED diagrams.

In Sections~\ref{secsymmetries} and \ref{secproperties}.1, we considered the C, P, T, CP, CT, PT, and CPT transformations of processes and diagrams. For each transformation, we gave the properties of the regularizing propagators and vertex function that suffice to conclude that:
\begin{itemize}
\item[(A)] There is a phase factor characteristic of this transformation, which equals a power of $-1$, and is common to all diagrams of the same process.
\item[(B)] The transition probability is the same for two processes related by this transformation.
\item[(C)] The transition probability vanishes for each process that involves a set of particles invariant under this transformation if the corresponding phase factor of diagrams equals $-1$. There are such processes for the C, P, T, and PT transformations; for C transformation, some of these results are implied by Furry's theorem.
\end{itemize}

Within the considered framework of regularized QED diagrams, there are three different reasons why a process is not possible and there is a selection rule:
\begin{itemize}
\item[1)] Each process has to conserve total electric charge or we cannot model it by the diagrams considered.
\item[2)] Transition probability vanishes for each process that does not abide by the Pauli exclusion principle, or does not conserve the total momentum or the total angular momentum.
\item[3)] Transition probability vanishes for each process that is invariant under a disrete symmetry if the corresponding phase factor of diagrams equals $-1$. This is true for any S-matrix, i.e.: If an S-matrix is invariant under a discrete symmetry up to a phase factor different from one, and the process is invariant under this symmetry, then the corresponding S-matrix element vanishes.
\end{itemize}

If there is a Lagrangian corresponding to the regularizing propagators and vertex function, one can incorporate the resulting transition amplitudes into a field-theoretic framework as defined by 't Hooft and Veltman\cite{Hooft}; for an example see~\cite{mi001}.

In Sections~\ref{secinvariances} and \ref{secproperties}.3 we introduced a new transformation of processes and diagrams, H, related to the Hermiticity of the propagators and vertex function. Combining H with C, P, and T transformations, we obtained seven new transformations. We gave properties of the regularizing propagators and vertex function that make the lowest-order transition probabilities the same for two processes related by these transformations.

In Section~\ref{secgeneral} we established the general, Lorentz form-invariant fermion and photon propagators and fermion-photon vertex function that make diagrams Lorentz-invariant. We gave conditions they have to satisfy so that the Lorentz-invariant diagrams are invariant up to a phase factor also under C, P, T, and H transformations and their products. These Lorentz-invariant diagrams are always invariant up to a phase factor under the CPT transformation---a special case of the CPT theorem.

\section*{Acknowledgement}

We are grateful to Matja\v z Polj\v sak for helpful conversations and suggestions.

\end{document}